\documentclass[elsart]{revtex4}

\usepackage{graphicx}

\begin{document}

\title{Dipole formation at metal/PTCDA interfaces: role of the Charge Neutrality
Level}
%\shorttitle{Dipole formation at metal/PTCDA interfaces: CNL}

\author{H. V\'azquez$^{\dagger}$, R. Oszwaldowski$^{\dagger}$,
 P. Pou$^{\dagger}$, J. Ortega$^{\dagger}$, R. P\'erez$^{\dagger}$, 
 F. Flores$^{\dagger}$ and A. Kahn$^{\P}$}
%\shortauthor{H. V\'azquez \etal}

%\affiliation{
\address{
 $^{\dagger}$ Departamento de F\'{\i}sica Te\'orica de la Materia Condensada-
Universidad Aut\'onoma de Madrid, E-28049 Madrid, Spain. \\
 $^{\P}$ Department of Electrical Engineering- Princeton University,
Princeton, NJ 08544, USA. }

%\pacs{71.20.Rv}{Electron density of states and band structure of crystalline
%solids. Polymers and organic compounds}
%\pacs{68.43.Bc}{Chemisorption/physisorption: adsorbates on surfaces. Ab-initio calculations of adsorbate structure and reactions}
%\pacs{73.30.+y}{Surface double layers, Schottky barriers and work functions}

%\date{\today}

%%%\begin{document}

%%%\maketitle

\begin{abstract}

The formation of a metal/PTCDA (3, 4, 9, 10- perylenetetracarboxylic
dianhydride) interface barrier is analyzed using weak chemisorption theory. The 
electronic structure of the uncoupled PTCDA molecule and of the metal surface is
calculated. Then, the induced density of interface states is obtained as a function of these two electronic structures and the interaction
between both systems. This induced density of states is found to be large enough 
(even if the metal/PTCDA interaction is weak) for the definition of a Charge 
Neutrality Level for PTCDA, located $2.45$ eV above the highest occupied
molecular orbital. We conclude that the metal/PTCDA interface molecular level
alignment is due to the electrostatic dipole created by the charge transfer 
between the two solids.

\end{abstract}

\maketitle

%\bigskip
%\narrowtext
%\section{Introduction}

Electronic materials made of molecular films are a fast developing
field, with many potential applications in organic-based devices. Designing 
new organic-based materials requires a detailed understanding of the different 
processes occurring in these devices. In particular, metal/organic and 
semiconductor/organic interface barriers play a decisive role \cite{Ishii,Shen}. However, the formation of barriers is not
yet well understood.

In the Schottky-Mott model of metal/organic interfaces, it is assumed that no 
interface dipole is formed at the junction, and that the position of molecular 
levels with respect to the metal Fermi level is defined by vacuum level
alignment. This situation
was disproved by Narioka {\em et al.} \cite{Narioka} who, using ultra-violet
photoemission spectroscopy (UPS), found large interface dipoles ($\sim
0.5 - 1.0 eV$) at several metal/organic interfaces. Independent data by Hill 
{\em et al.} \cite{Hill} confirmed this conclusion. Various mechanisms are 
believed to operate simultaneously at these interfaces, and several models have
been advanced \cite{Ishii,Shen}. Metal-molecule chemical reaction has been seen
to create interface gap states that pin the Fermi level \cite{Shen2}, a
situation that is analogous to that described by the Unified Defect Model
proposed for inorganic semiconductor/metal interfaces \cite{Spicer}. Compression of
the metal surface electronic tail by adsorbed molecules, leading to vacuum level
interface shift (the ``pillow" effect), has also been proposed as a general 
metal/organic interface mechanism \cite{Ishii2,Crispin,Koch}.

In this letter, we explore the first application to a metal/organic
interface of the Induced Density of Interface (or virtual)
States (IDIS) Model \cite{Flores}. We study a metal/PTCDA (3, 4, 9, 10- perylenetetracarboxylic
dianhydride) interface and analyze how the chemical interaction 
between the organic molecule and the metal creates an IDIS in the organic
energy gap. Our calculations show that, although the chemical interaction is 
weak, the IDIS is large enough that a Charge Neutrality Level
(CNL) of the organic molecule can be defined. Our results show that
the interface Fermi level $E_{F}$ is pinned at the CNL, a situation similar to
that described for the formation of Schottky barriers at conventional 
semiconductor/metal junctions.

%\section{Molecular electronic structure}
\bigskip

In this theoretical analysis, we study the metal/PTCDA
interaction in three steps. First, the molecular orbitals of the organic molecule are calculated using a Density
Functional Theory (DFT) local-orbital method \cite{Pou,Rafal}.
Care is taken to identify the appropriate
wavefunctions associated with the Affinity and Ionization levels defining the
molecular transport gap. Then, we calculate the metal/molecule
interactions and obtain the electronic induced density of states 
in the molecular
energy gap. In this approach, we neglect the molecule-molecule
interaction, which introduces only a small broadening of the molecular
levels ($\sim 0.2$ eV), and does not create any electron density of
states in the molecular energy gap. Finally, we define the CNL 
associated with these induced density of states and show
that the organic/metal interaction creates 
density of states large enough to pin the
Fermi level close to the CNL. Pinning can be characterized by the parameter 

\begin{equation}
S = \frac{d E_{F} } {d \phi_{M} },
\end{equation}

(where $\phi_{M}$ is the metal workfunction), equivalent to the 
interface slope parameter in the theory of metal/semiconductor 
junctions \cite{Sze} and which we find here to be $\sim 0.13$.

\bigskip

In our DFT study of the PTCDA molecule \cite{Rafal}, 
we use an optimized minimal local
orbital basis, and introduce many-body effects by means of a local orbital 
exchange and correlation potential \cite{Pou}. In our actual 
calculations, we combined a local-orbital DFT code \cite{Fireball} with this
exchange and correlation potential. This approach is equivalent
to conventional DFT but is especially advantageous in cases where correlation
is important \cite{IJQC}, and for calculating the Affinity ($A$) and Ionization
($I$) levels of the molecule, as will be seen below in the discussion of
Koopman's theorem \cite{Rafal}.

\begin{figure}
%\onefigure[scale=0.35]{./Fig1.eps}
\includegraphics[scale=0.35]{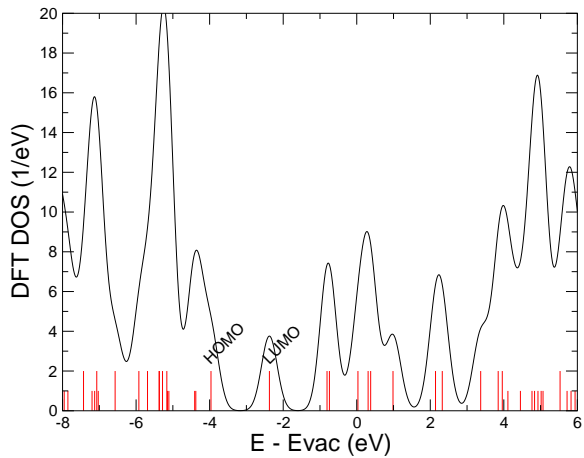}
\caption{
DFT spectrum (long bars: $\pi$ states; short bars:
$\sigma$ states) and DOS (introducing a 0.5 eV FWHM Gaussian broadening)
for the PTCDA molecule.
}
\label{Fig1}
\end{figure}

%\bigskip

Figure \ref{Fig1} shows the spectrum obtained from the DFT calculation; 
we also calculate a LDOS by introducing
for each level a Gaussian broadening with a $0.5$ eV FWHM \cite{Hill2}. This 
DFT calculation produces a molecular energy gap $\sim 1.6$ eV, very small 
compared to the theoretical single-particle gap $A - I$ ($\sim 5$eV) of the isolated
molecule \cite{polacos}, and even smaller than its optical gap, 
$\sim 2.6$eV \cite{Shen}. 
This is due to the fact that the DFT eigenvalues are not 
directly related to the molecular levels. Note also that, in analyzing the
metal/organic interface, one needs to consider the single-particle energy gap of 
the isolated molecule, $A-I$. At the interface, this energy gap, $A-I$, 
is reduced by solid-state effects \cite{Shen,Hill2,Tsiper}, as will be 
discussed later. 

%\section{Transport energy gap}
\bigskip

%Let us now consider how to calculate the single-particle gap, $A-I$, for the 
%isolated molecule. 
We first calculate $A$ and $I$ using the definitions
$A = E[N+1] - E[N]$, and $I = E[N] - E[N-1]$, where $E[N_{i}]$ is the ground-state
energy of the molecule with $N_{i}$ electrons. Each case can be
calculated using the DFT method described above; we find \cite{Rafal}
$I=-5.9$ eV and $A=-1.1$ eV, yielding an energy gap of $4.8$ eV, in reasonable
agreement with the value of $\sim 5$ eV \cite{polacos}. 
Interestingly, similar results can be obtained using a sort of 
Koopman's theorem \cite{Rafal}, 
which amounts to introducing an electron (hole) in 
the LUMO (HOMO) of the neutral 
molecule, and calculating the total energy of the system 
{\em neglecting electron relaxation effects}. 
In this approach, the Affinity level, for 
instance, is simply the LUMO level corrected by a quantity $\delta A$
\cite{Rafal}, which can be interpreted as 
a self-interaction energy associated with the new electron charge at the
LUMO wavefunction. Likewise, the ionization-level correction, $\delta
I$ \cite{Rafal}, is a negative self-interaction energy related to the
removal of one electron from the HOMO.

This Koopman's theorem approach yields $I=-6.1 eV$ and $A=-1.2 eV$, 
in very good 
agreement with the values
given above. This also shows that electron relaxation effects 
in PTCDA associated with
introducing or extracting an electron from the molecule are small, and
indicates that the molecular orbitals calculated within DFT for the HOMO and
LUMO levels are very appropriate for describing the Ionization and Affinity
wavefunctions of the molecule. Note that $\pi$ and
$\sigma$ orbitals give different values of $\delta A$ and $\delta I$. In our
calculations, we obtain the PTCDA energy levels for the molecule by modifying the
DFT $\pi$ and $\sigma$ levels with the corresponding values of $\delta A$ and $\delta
I$ given by Koopman's theorem.

Solid-state effects associated with long-range electronic polarization have
been analyzed by Tsiper {\em et al.} \cite{Tsiper}. Polarization of molecules of 
the solid screens the electric field created by the
extra charge introduced in the system, either an electron or a hole, modifying
the energy levels $I$ and $A$ and the corresponding single-particle energy gap
(also called transport gap in the solid). This
correction is important, $1 - 1.5 eV$, yet the ionization and affinity
wavefunctions of the molecule are not expected to present important
modifications. Other effects, such as lattice relaxation and vibronic
coupling, only introduce corrections of the order of $0.2 eV$, in the transport
gap \cite{Hill2}.

Thus, the electronic wavefunctions
of PTCDA can be described using the results of a DFT approach, 
introducing a Koopman's correction for the $\pi$ and $\sigma$ molecular energy
levels. Polarization effects are then included by fitting the ionization level
and transport gap to their experimental values \cite{Transpgap} by a rigid shift in
the occupied and empty electron states. We fit the energy gap, $A - I$, 
to $3.2$ eV, which seems to include all the effects discussed above.

%\section{DOS and CNL}
\bigskip

We analyze the metal/PTCDA contact using chemisorption theory in the
limit of weak interaction. 
The inset of Figure \ref{Fig3} shows the geometry
of interest: a planar PTCDA molecule located at a distance $d$ from
the outermost metal layer, i.e. the Au $(111)$ surface. 
We assume the PTCDA wavefunctions, $\psi_{i}$, to be described by
the DFT method discussed above, and the metal wavefunctions by a DFT-LDA
local-orbital code \cite{Fireball}, which also yields the Green-function, 
$G_{\alpha \beta}(E)$, where $\alpha$ and $\beta$ refer to the local-orbital
basis used in the DFT code. In the limit of weak metal-PTCDA interacion \cite{Hill2},
the main effect of the metal on the PTCDA electronic structure is to 
introduce the following self-energy for the
molecular levels:

\begin{equation}
\Sigma_{i} (E) = \sum_{\nu} \; |T_{i\nu}|^{2} \; G_{\nu}(E) \label{self}
\end{equation}

where $T_{i\nu}$ is the hopping interaction between the molecular orbital 
$\psi_{i}$ and the metal eigenfunction, $\psi_{\nu}$,
$G_{\nu}(E)$ being the corresponding Green-function. 
Equation \ref{self} can
be rewritten in a more appropriate way by using a local-orbital basis for the
molecule and the metal. With the notation 
$\psi_{i} = \sum_{j} c_{ij} \phi_{j}$, equation \ref{self} takes the form

\begin{equation}
\Sigma_{i}(E) = \sum_{j j^{\prime} \alpha \beta} c_{ij} \; T_{j\alpha} \; 
G_{\alpha\beta} (E) \; T_{\beta j ^{\prime}} \; c_{j ^{\prime} i},
\label{sigmaicij}
\end{equation}

where the self-energy of each molecular level is determined by: 
$(a)$  the hopping
interaction $T_{j\alpha}$ between local orbitals; $(b)$ the coefficients
$c_{ij}$ of the molecular orbitals $\psi_{i}$;
and $(c)$ the Green-function 
of the metal, $G_{\alpha \beta}(E)$.

\begin{figure}
%\onefigure[scale=0.35]{./Fig2.eps}
\includegraphics[scale=0.35]{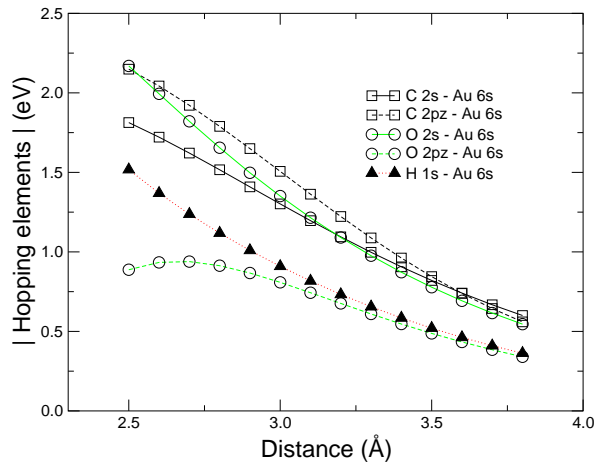}
\caption{
C, O and H - Au interaction as a function of PTCDA-Au distance.}
\label{Fig2}
\end{figure}

\begin{figure}
%\onefigure[scale=0.35]{./Fig3.eps}
\includegraphics[scale=0.35]{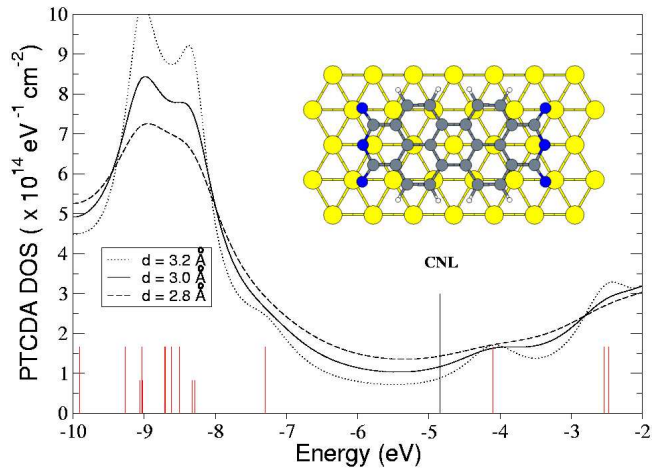}
\caption{
PTCDA LDOS for the PTCDA/Au$(111)$ interface (top view at inset) calculated for
$d=2.8$ {\AA} (dashed line), $d=3.0$ {\AA} (continuous line), and $d=3.2$ {\AA} 
(dotted line). Long (short) bars correspond to the $\pi$ ($\sigma$) states 
neglecting the metal-molecule interaction.
}
\label{Fig3}
\end{figure}

Figure \ref{Fig2} shows the hopping interaction
between the Au $6s$ orbital and the different orbitals of PTCDA, C $2s$ and $2p$,
O $2s$ and $2p$, and H $1s$ (in the following, we neglect the $5d$ and $6p$
orbitals of Au, which is a reasonable approximation because we are 
interested in the induced density of states around the metal Fermi energy). 
These hopping elements are calculated using DFT for each pair
of molecule-substrate atoms.

\bigskip

Equation \ref{sigmaicij} can be simplified to a more convenient form by
neglecting contributions having $j \neq j^{\prime}$: these terms represent
interference effects between different local orbitals of PTCDA, and tend to
cancel each other out. The contribution coming from the Au $6s$ orbitals yields

\begin{equation}
\Sigma_{i}(E) = \sum_{j,6s} |c_{i,j}|^{2}  |T_{j,6s}|^{2} 
G_{6s} (E).
\label{sigmai6s}
\end{equation}

%\begin{equation}
%- \frac {1} {\pi} Im \left\{ \frac {1} {(E-E_{i}) - \Sigma_{i}(E)}
%  \right\}.
%\end{equation}
Each molecular level contributes to the LDOS with $- \frac {1} {\pi}$
Im $\left\{ \frac {1} {(E-E_{i}) - \Sigma_{i}(E)} \right\}$.
The main effect of $\Sigma_i$ is to broaden the molecular levels by 
$Im ( \Sigma_i (E))$. Since we are studying the
density of states around the molecular energy gap, we take 
$\Gamma_i/2 \simeq \Sigma (E_F)$ ($E_F$ is the interface Fermi 
level), and obtain

\begin{equation}
\frac {1} {\pi} \frac {\Gamma_{i}/2} {(E-E_{i})^{2} + {(\frac{\Gamma_{i}}
{2})^{2}}}
\end{equation}
for the contribution of each molecular level to the LDOS.

As shown in Figure \ref{Fig2}, $T_{j,6s}$ and $\Gamma_i$ depend on the PTCDA-Au
distance, $d$. We do not know of any experimental value for this
interfacial distance, nor are DFT calculations reliable in obtaining it,
since the metal/PTCDA interaction has an important Van der
Waals (or dispersion forces) contribution. Recently, Krause {\it et al.} 
\cite{Krause} measured the PTCDA-Ag(111) distance to be $\sim 2.85 $ \AA,
with an uncertainty of $\pm 0.2$ \AA. We expect a similar (or slightly larger)
distance, of $\sim 3.0$  \AA \ for the PTCDA-Au(111) system.

Figure \ref{Fig3} shows the PTCDA LDOS (both spins included) around the organic 
energy gap for $d= 2.8, 3.0$ and $3.2$ \AA. For these distances, the $\pi$ and 
$\sigma$ levels are broadened by 
$\Gamma^{\pi}_i \simeq 2.1, 1.5$ and $1.0$ eV and 
$\Gamma^{\sigma}_i \simeq 1.05, 0.75$ and $0.5$ eV, respectively.
The interactions $T_{j,6s}$ and the level broadening terms
$\Gamma^{\pi}_i$ and $\Gamma^{\sigma}_i$ increase with decreasing
distance: for $d= 2.8 $ \AA \ the values of $\Gamma$ are typically
twice the values for $d=3.2 $ \AA. Figure \ref{Fig3} also shows the position of 
the CNL. The CNL is calculated by integrating the induced LDOS and imposing
charge neutrality conditions: the total number of electrons up to the CNL
equals that of the isolated molecule. The CNL is practically 
the same for the three distances $d= 2.8, 3.0$ and $3.2$ \AA, showing that it 
is a very robust quantity,
practically independent of the details of the metal/PTCDA interaction.

We calculated the interface slope parameter,
$S = \frac {d E_{F}} {d \phi_{M}}$. In our 
case \cite{Sze},

\begin{equation}
S = \frac{d E_{F} } {d \phi_{M}} =  \frac {1} {1 + 4 \pi e^{2} D(E_{F}) 
\delta / A },
\end{equation}

where $D(E_{F})$ is the PTCDA LDOS (Figure \ref{Fig3}) at the Fermi level 
(practically, the CNL), $\delta$ is the
distance between the charges induced in the metal and the organic molecule
(between $2.8-3.2$ \AA \ in Figure \ref{Fig3}) 
and A is the area associated with a PTCDA molecule (around
120 \AA$^{2}$). We find a DOS of $1.5, 1.2$ and $0.9 \times 10^{14}$ eV$^{-1}$
cm$^{-2}$ for $d= 2.8, 3.0$ and $3.2 $ 
\AA, respectively. For these distances, $S \simeq 0.12, 0.13$ and $0.16$. 
At variance with the CNL, $S$ strongly 
depends on the metal-PTCDA distance. But even for $d = 3.5$ \AA, 
we obtain a DOS of $0.5 \times 10^{14}$ eV$^{-1}$ cm$^{-2}$ and S
$\simeq 0.23$, indicating that the IDIS is still significant.
This small value of S ($0.1-0.2$) 
shows that the LDOS induced in the molecule is high enough
(even if the metal/PTCDA interaction is small) to strongly pin the interface
Fermi energy close to the CNL. 

Up to this point we have assumed the molecule to present no geometrical
distortion when deposited flat on the metal. Recent DFT calculations for organic 
molecules adsorbed on metals \cite{ostrom} show non-negligible bond-length 
changes ({\it e.g.} the C-C bond in an octane molecule adsorbed on Cu(110) is 
shortened by $\sim 0.04$ \AA). However, given the linear
geometry of the alkanes, their geometrical distortions are
probably larger than the ones in PTCDA on Au$(111)$. Nevertheless, we
explored the sensitivity of our results to geometrical distortions 
by changing some of the molecular bond lengths. We analyzed how the
deformation of the molecular C-C and C-O bonds modify the position of the CNL
and the LDOS. Our calculations indicate that the CNL shifts
by 0.1 eV for a change of 0.1 \AA \ of the carboxyl C-C bonds and by $\sim 0.07$ 
eV for a 0.1 \AA \ change in the C-O bonds, 
while in all cases the LDOS around the CNL is practically unaltered.
Thus, we conclude that the CNL undergoes only minor changes due to the 
molecular deformation, and that the LDOS is not significantly modified.
The CNL shift due to molecular distortions
can be estimated by realizing that the weaker single C-C bonds are probably 
going to suffer the largest deformation. In the case of the alkanes,
these distortions are $\sim 0.05$ \AA \cite{ostrom}.
This suggests that the accuracy in our calculated CNL, neglecting molecular
deformations, is better than $\pm 0.1$ eV.

From our calculations and the previous discussion, we find
that the CNL is 
located $2.45 \pm 0.1$ eV above the center of the HOMO level of the molecule, in very
good agreement with the experimental $E_{F}$ pinning position \cite{Hill}. 
The value of $S \simeq 0.13$ is also in good agreement with $S \simeq 0$ 
inferred from experiments \cite{Hill}. The important
outcome of our analysis is that the calculated value of $S \simeq 0.13$ 
is indicative of a
significant DOS at the metal/PTCDA interface, in spite of the weak
chemical
interaction between the two materials (see Figure \ref{Fig2}). The implication 
is that the interface Fermi level is close to the CNL (in our calculations,
$E_{F}$ lies $\sim 0.02$ eV below the CNL). 
Thus, the
formation of the interface barrier is related to the transfer 
of charge across the interface. The charge transfer is associated with the 
IDIS, and creates an electrostatic interface dipole, which tends 
to align the metal Fermi level and the
PTCDA CNL.

Recently, Bagus {\it et al.} \cite{Bagus} have suggested that for
molecules physisorbed on metal surfaces a significant interface dipole,
$\sim 0.3$ eV, originates from exchange-like effects
(this is the ``pillow" effect mentioned above
\cite{Ishii2,Crispin,Koch}).
This effect can be easily incorporated in our discussion as a
modification of the initial metal workfunction. As the final interface
Fermi level $E_F$ is given by:

\begin{equation}
(E_F - CNL) = S ( \phi_M - CNL ) ,
\end{equation}
where $E_F$, CNL and $\phi_M$ must have a common origin, a change of 0.3 eV in 
$\phi_{M}$ is reflected in a Fermi level shift 
of 0.04 eV ($S = 0.13$). This suggests that for PTCDA, where $S$ is small,
the ``pillow" effect is almost negligible with regard to the final Fermi
level position.

%\section{Conclusions}
\bigskip

In conclusion, we have analyzed the metal/PTCDA interface barrier formation 
using weak chemisorption theory. The aim of this paper was to explore the 
importance of the IDIS and its contribution to the interface formation.
Our results show that even for Au-PTCDA distances for which the
interaction is weak ($d \sim  3.0 - 3.5 $ {\AA}), 
the IDIS is high enough to create an 
interface dipole that tends to align the Fermi level and the PTCDA CNL. A
measure of this drive to align the two levels is given by the slope parameter, 
$S$, which our calculations show to be $\sim 0.13$.

Within this framework, we conclude that the main mechanism that leads to
the formation of the Schottky barrier in metal/PTCDA
interfaces is the charge transfer associated with the tunneling of
metal electrons through the molecular energy gap. Our results show that the
interface Fermi level should be close to the CNL of the molecule, which our
calculations place $2.45 \pm 0.1$ eV above the HOMO level.

\acknowledgements
We gratefully acknowledge financial support by the Spanish CICYT under project 
MAT 2001-0665, the Comunidad de Madrid and the DIODE network (HPRN-CT-1999-00164). 
Support of this work by the National Science Foundation (DMR-0097133) and the
New Jersey Center for Organic Optoelectronics (AK) is also acknowledged.

\end{document}